# An infectious disease model on empirical networks of human contact: bridging the gap between dynamic network data and contact matrices

A. Machens[1,2,3], F. Gesualdo[4], C. Rizzo[5], A.E. Tozzi[4], A. Barrat[1,2,3,*], C. Cattuto[3]

[1]Aix Marseille Université, CNRS UMR 7332, CPT, 13288 Marseille, France
[2]Université du Sud Toulon-Var, CNRS UMR 7332, CPT, 83957 La Garde, France
[3]Data Science Laboratory, ISI Foundation, Torino, Italy
[4]Bambino Gesù Children's Hospital, IRCCS, Rome, Italy
[5]National Centre for Epidemiology, Surveillance and Health Promotion, Istituto Superiore di Sanità Rome, Italy

[*]Corresponding author: A. Barrat, alain.barrat@cpt.univ-mrs.fr

**Abstract**

Background
The integration of empirical data in computational frameworks designed to model the spread of infectious diseases poses a number of challenges that are becoming more pressing with the increasing availability of high-resolution information on human mobility and contacts. This deluge of data has the potential to revolutionize the computational efforts aimed at simulating scenarios, designing containment strategies, and evaluating outcomes. However, the integration of highly detailed data sources yields models that are less transparent and general in their applicability. Hence, given a specific disease model, it is crucial to assess which representations of the raw data work best to inform the model, striking a balance between simplicity and detail.

Methods
We consider high-resolution data on the face-to-face interactions of individuals in a pediatric hospital ward, obtained by using wearable proximity sensors. We simulate the spread of a disease in this community by using an SEIR model on top of different mathematical representations of the empirical contact patterns. At the most detailed level, we take into account all contacts between individuals and their exact timing and order. Then, we build a hierarchy of coarse-grained representations of the contact patterns that preserve only partially the temporal and structural information available in the data. We compare the dynamics of the SEIR model across these representations.

Results
We show that a contact matrix that only contains average contact durations between role classes fails to reproduce the size of the epidemic obtained using the high-resolution contact data and also fails to identify the most at-risk classes. We introduce a contact matrix of probability distributions that takes into account the heterogeneity of contact durations between (and within) classes of individuals, and we show that, in the case study presented, this representation yields a good approximation of the epidemic spreading properties obtained by using the high-resolution data.

Conclusions
Our results mark a first step towards the definition of synopses of high-resolution dynamic contact networks, providing a compact representation of contact patterns that can correctly inform computational models designed to discover risk groups and evaluate containment policies. We show in a typical case of a structured population that this novel kind of representation can preserve in simulation quantitative features of the epidemics that are crucial for their study and management.

**Background**

Computational models and multi-scale numerical simulations represent essential tools in the understanding of the epidemic spread of infections, in particular to study scenarios, and to design, evaluate and compare containment strategies. The applicability of such computational studies crucially depends on informing transmission models with actual data. We are currently witnessing an important evolution as more and more data on human mobility and behavioral patterns become accessible (1, 2, 3, 4, 5, 6, 7, 8, 9). For instance, data on human travel patterns and mobility have been fed into large-scale models of epidemic spread at regional or planetary scale, providing important modeling and prediction tools (10, 11, 12, 13, 14).

When considering smaller scales, the knowledge of contact patterns among individuals becomes relevant for identifying transmission routes, recognizing specific transmission mechanisms, and targeting groups of individuals at risk with appropriate prevention strategies or interventions such as prophylaxis or vaccination (15). Several properties of the contact patterns are known to bear a strong influence on spreading patterns, such as the topological structure of the contact network, the presence of individuals with a particularly large number of contacts, the frequency and duration of contacts, and the existence of communities (16, 17, 18, 19, 20, 21, 22, 23, 24).

The simplest approach to the description of contact patterns is the homogeneous mixing assumption, which postulates that each individual has an equal probability of having contacts with any other individual in the population (25). A widely used refinement of this approach consists in dividing individuals into classes (corresponding for instance to different age groups or to different social activities), and defining a contact matrix between classes in terms of the average number (or duration) of the contacts that individuals in one given class have with individuals in another given class. These matrices are constructed using data from questionnaires or diaries, time-use data (23, 26, 27, 28, 29, 30, 31), and more recently by sensing room co-presence and close-range proximity between individuals (32, 33).

The use of models with structured populations can help to define more refined analytical approaches (34). In addition, it allows to perform numerical simulations of epidemic spread in synthetic structured populations with fixed rates of contacts between individuals belonging to different classes, given by the (possibly empirical) contact matrix elements (12, 10, 35, 36, 37). These simulations can also be used to compare vaccination strategies targeting specific groups and to estimate which strategies are most effective (35, 36, 37, 38, 39, 40, 41). The use of contact matrices for modeling contact patterns relies on a set of restricted homogeneous mixing assumptions within each class and on the representativeness of the average mixing behavior between classes. However, such an approach neglects the strong fluctuations that are usually observed in the distributions of the numbers and durations of contacts between two individuals of given classes, which are often modeled with negative binomial distributions (29, 23, 31, 42, 43).

Overall, in the context of a specific modeling problem, little is known about the level of detail that should be incorporated in modeling contact patterns. Coarse representations such as the homogeneous mixing assumption leave out crucial elements, but are analytically tractable and can provide a coarse understanding of epidemic processes. More realistic approaches are however needed when the aim is to predict with quantitative accuracy the outcomes of specific scenarios, to target groups of individuals who are most at risk, and to quantitatively evaluate and rank interventions and containment strategies according to their effectiveness. In these cases it is crucial to achieve realistic simulations to estimate quantities such as the extinction probability or the attack rate. However, very detailed representations may critically lack transparency about the specific role of the many modeling assumptions they incorporate, and might yield unnecessarily fine-grained predictions. For instance, in order to evaluate the relative efficacy of targeted vaccination of different groups of persons, we do not need to know the risk of infection of each individual — it is sufficient to estimate the average attack rate in each group.

As technological advances keep enhancing our ability to gather high-resolution data on proximity and face-to-face interactions (4, 5, 6, 7, 8, 9) and to integrate such data in computational models, the issue of understanding what is the most adequate data representation becomes therefore crucial, and the answer may depend on the specific epidemic process under study as well as on the specific goals of the modeling approach. This points to general problems for data-rich scenarios such as the ones involving wearable sensors in real-world settings: What is the right amount of information about individual interactions that is appropriate for a given modeling task, so that the relevant information is retained but the model stays as parsimonious as possible (44)? What are the most useful synopses of high-resolution contact network data?

Several studies have started to tackle such issues within the framework of the description of human interactions as static or dynamic contact networks, in the case of unstructured populations (22, 45, 46 24, 47, 48). It is known that the heterogeneity of contact durations is a crucial element that needs to be taken into account. On the other hand, in a particular case (48) it was shown that the dynamics of an SEIR (Susceptible, Exposed, Infectious, Recovered) process over an aggregated network that only takes into account daily contact durations achieves a good approximation of the transmission dynamics over the full time-varying contact network with a temporal resolution of the order of minutes.

To date, the more complex case of a structured population has not been investigated under this perspective of data summarization. Here we address this problem motivated by the needs of applications and interventions aimed at the containment of diseases, such as identifying groups of individuals that need to be prioritized in deploying containment strategies such as vaccination and prophylaxis.

The aim of the present paper is therefore to understand whether and how different representations of the contact patterns within a structured population might lead, in simulation, to different estimates for the outcomes of an epidemic process and for the attack rates within groups of individuals: How faithful to the high-resolution empirical data are the summarized representations of contact patterns, in terms of the spreading process they yield and of the risk groups they identify? Given a specific epidemic process to be modeled, what is the optimal tradeoff between the compactness of the data representation and the amount of detail retained in the model? The answer to these questions lies in designing compact yet informative representations of contact patterns that can be used to model epidemics in structured communities (e.g., the case of nosocomial infections in hospitals). In such contexts, the design and evaluation of prevention measures or containment policies is naturally based on targeting specific classes of individuals. For example, it is possible to prioritize the vaccination of a certain class of individuals (e.g., nurses in a hospital (35)), whereas no simple policy can be based on a list of specific individuals. Therefore, it is important to build data representations that yield accurate results at the level of groups of individuals and that, at the same time, can be generalized in order to support generic conclusions on the relative efficacy of different strategies, and in particular of strategies targeting only high risk groups.

Here we put forward such a representation and we validate its appropriateness and usefulness in a case study of structured population: to this aim, we leverage a high-resolution dataset on the face-to-face proximity of individuals in a hospital ward, collected by the SocioPatterns collaboration by using wireless wearable sensors. The population is structured in different classes that correspond to different roles in the hospital ward. The collected data, described in detail in Ref. (33), include the detailed time-ordered sequence of individuals' contacts between all subject pairs.

We consider various representations of the hospital ward data, corresponding to different types and degrees of aggregation. At the most detailed level, the raw data can be viewed as a dynamic contact network, where all the available information on the interactions between pairs of individuals (contact times and durations) is explicit. We then construct static networks that project away the temporal structure of the contacts but retain the identity of each individual and the structure of her contact network. At a coarser level, we consider the customary contact matrix representation of the data, which aggregates along the class dimension: All individuals belonging to the same role class are grouped together, and the contact rate between two individuals of given classes is given by the corresponding element of the contact matrix. This representation, by definition, discards many heterogeneities of contact behavior at the individual level as well as the heterogeneities among contacts. Between the contact matrix view and the individual-based network view, other intermediate representations are possible. Here we introduce a novel representation based on a contact matrix of distributions, designed to retain the heterogeneous properties of contact durations among pairs of individuals belonging to given role classes.

We then use an SEIR process to model the spread of an infectious disease in a structured hospital population whose contact patterns are described by using all of the above representations, computed from the raw proximity data. We simulate the distribution of attack rates for the various role classes and compare these outcomes against individual-based simulations that take into account the most detailed description of the empirical contacts. Our results highlight similarities and differences in the dynamics obtained for various aggregation levels of the contact pattern representation. In particular, they show that coarse representations that do not take into account the heterogeneity of contact durations, such as the usual contact matrix, yield rather bad estimates for the global attack rate. Such representations also lead to a wrong classification of role classes in terms of their relative risk and could therefore lead to incorrect prioritization for interventions such as targeted vaccination or prophylaxis. The novel representation that we put forward is shown to strike a promising balance between simplicity and generalizability on the one hand, and on the other hand the ability to yield accurate evaluations of risk and a correct prioritization of groups of individuals in the design of targeted measures.

**Methods**

**Data**

The dataset we use (33) describes the contacts of 119 individuals in a general pediatric ward of the Bambino Gesù Children's Hospital in Rome, Italy, over a period of one week (9-16 November 2009). The ward under study is located in the Department of Pediatrics and is physically separated from other wards and facilities of the Hospital. It has 44 beds arranged in 22 rooms with 2 beds each and mostly admits children with acute diseases who do not require intensive care or surgery. Admitted patients are accompanied by one person (parent or tutor) who spends the night in the same room. Contacts, defined as close proximity interactions, were measured by using wearable badges that engage in bidirectional low-power radio communication (7, 8) and can assess the face-to-face proximity of two individuals.

The study was approved by the Ethical Committee of the Bambino Gesù Hospital (33). As described in (33), during the data collection period all healthcare workers (HCWs) on duty, patients, parents or tutors were invited to participate in the measurements. HCWs participated in several meetings to receive information about the study and all procedures were approved by the Ethical Committee of the hospital. Only 5 HCWs, 1 patient and 1 parent declined. Some proximity sensing badges assigned to patients, parents or tutors were not continuously detected by the receiving infrastructure and were missing for more than 25% of the time they were assigned to the individuals. These badges were excluded from the dataset, after assessment of the robustness of the data statistical properties with respect to this filtering procedure (see (33) for details). Overall, the data describes the contacts of 119 individuals, among which more than 90% of the HCWs who were on duty in the ward during the chosen week. These 119 participants were associated with the following role classes: 37 patients (P), 20 physicians (D), 21 nurses (N), 10 ward assistants (A), i.e., health care workers in charge of cleaning the ward and distributing the meals, and 31 caregivers (C), who comprise tutors and non-professional visitors. As data was collected during the pandemic period when several patients with H1N1 infection were hospitalized, HCWs applied standard operating procedures to prevent transmission of infection including hand hygiene, face mask, wearing gowns and gloves. Specific guidelines are however permanently adopted in the study hospital to prevent cross transmission of respiratory infections, so that the contact patterns were not specifically altered during the data collection period.

**Empirical contact patterns**

An analysis of the recorded contact patterns is described in Ref. (33). At the most detailed level the data provide an explicit representation of the time-varying contact network of individuals, resolved at the fastest available temporal scale (20 seconds). As discussed in the Introduction, our aim here is to compare different numerical simulations of the spread of an infectious disease, where each simulation is constructed on top of a specific mathematical representation of contact patterns, and all these representations are derived from the *same empirical data*, summarized or modeled at different levels of detail (e.g., individual-based contact network vs contact matrices). In the following we describe the contact pattern representations we will use.

The representation that most closely corresponds to the raw empirical data, here forth named "DYN", is the time-varying individual-based contact network, where for each pair of individuals the precise starting and ending times of individual contact events are available with the finest temporal resolution of 20 seconds. When simulating a process, such as the spread of an infectious disease, characterized by longer time scales than the available interval of one week, we simply play back the recorded contact history over and over. Whereas other procedures are possible for artificially extending the time span of the empirical data, it was shown in (48) that the simple repetition procedure gives robust results.
The DYN representation is thus the one that contains the most detailed information on the empirical contact patterns, and therefore we will consider it as the reference against which all the other representations have to be compared.

Starting from the dynamical contact network representation (DYN), different aggregation procedures are possible that retain only selected features of the contact network while aggregating over the others. At the most aggregated level we consider, we disregard the topology of the empirical contact network as well as the contact heterogeneities and assume that each individual is in contact with all others for the same average time. We refer to this very coarse representation as "FULL", because its contact structure is a fully connected graph, with homogeneous contact durations.

As pointed out above, several intermediate representations can be designed between the fine-grained DYN representation and the coarse-grained FULL one. We define a space of possible representations by considering two qualitative features: the amount of information on the network structure, and the amount of information on contacts (times, durations, etc.). Figure 1 and Table 1 illustrate where the representations we study, described in the rest of this section, lie with respect to one another, and which properties of the empirical data they preserve.

The first two representations we consider are designed to preserve the empirical structure of the contact network (who has met whom). The first representation (HET in Fig. 1) is a *static* and *weighted* contact network: individuals who have been in contact during the data collection are linked by an edge that is weighted by the total time they have spent in face-to-face proximity. All the dynamical contact events are thus summarized into a single contact graph that includes information on the observed contacts between individuals (who has met whom) and on the total (heterogeneous) duration of these contacts, i.e., how long A and B have been in contact. HET disregards information on the time of occurrence and on the temporal ordering of contacts. A less informative static network representation can be obtained from HET by removing the weight heterogeneity: this homogeneous network (HOM) is constructed by retaining the topology of the HET network, while giving all links the same weight, set as the average total duration of contacts between two individuals that have met at least once.

The DYN, HET and HOM representations of the raw data describing the contacts between the attendees of a conference were compared in Ref. (48). In the present case of a hospital ward, however, role classes are present, and the structure of the contact network is known to be very different (33) because of the role structure. It is therefore interesting to first verify the validity of the results of Ref. (48) in the case of a structured population.

The role structure of a hospital ward naturally lends itself to the definition of *contact matrices* that encode in a very compact way the class heterogeneity of contact patterns. Rows and columns of a contact matrix correspond to role classes. The matrix entry for classes X and Y gives the average contact time between any two members of those respective classes. The contact matrix description is not individual-based, hence in order to make contact with the individual-based representations introduced above, as well as with the empirical data, we define a network representation of the contact matrix (CM) as follows. We build a complete contact graph where each pair of individuals (x,y), with x in class X and y in class Y, is linked by an edge of weight $w_{XY}$ that depends only on the respective classes of the individuals. We set $w_{XY} = W_{XY}/(N_X N_Y)$, where $N_X$ is the number of individuals in class X, $N_Y$ the number of individuals in class Y, and $W_{XY}$ is the total time spent in contact by any individual of class X with any individual of class Y. In the case X=Y, we set $w_{XX} = W_{XX}/(N_X(N_X-1)/2)$, where the denominator is the number of within-class edges. Similarly to the FULL representation, the CM representation does not preserve the topology of the empirical contact network, but the more coarse-grained contact structure between classes of individuals is retained by associating pairs of individuals with weights that depend on their classes.

As already mentioned above, it is known (33) that the distribution of cumulative contact durations between pairs of individuals belonging to given categories displays large heterogeneities. In fact, as reported in other studies (29, 23, 31. 42, 43), such distributions can often be modeled as negative binomial distributions, which can account for the broad fluctuations of contact durations as well as for the comparatively high fraction of missing links across the chosen classes. In order to account for these heterogeneities, we define a new representation based on a contact matrix of distributions. The matrix is defined so that the entry for role classes X and Y consists of the parameters of the negative binomial distribution that fits the empirical distribution of contact durations between all pairs of individuals x in class X and y in class Y.

Similarly to the customary contact matrices, the contact matrix of distributions is not an individual-based representation. Hence, in analogy to what we have done above, we start from the *Contact Matrix of Distributions* and define a corresponding network representation ("CMD", in the following): we build a complete contact graph, where each pair of individuals *(x,y)*, with *x* in class X and *y* in class Y, is linked by an edge whose weight $w_{xy}$ is a random variable obtained by sampling the negative binomial distribution with the parameters specified by the (X,Y) entry of the contact matrix of distributions. A weight $w_{xy}=0$ indicates that there is no link between individuals *x* and *y*. We remark that the detailed structure of the empirical contact network is not retained in this representation, but that the average number of links with non-zero weight between each pair of classes is preserved.

Finally, we also report on an intermediate contact matrix representation designed to capture the link density and the average weights of the contact sub-network connecting classes X and Y, without taking into account the corresponding variance of edge weights. This is built as a contact matrix of bimodal distributions (CMB, "B" for "bimodal") that are used in place of the negative binomial distributions. As in the CMD case, the matrix is defined so that the entry for role classes X and Y consists of the parameters of a bimodal distribution that summarizes the empirical contacts between all pairs of individuals x in class X and y in class Y. Specifically, let us denote by $E_{XY}$ the number of observed edges between individuals of classes X and Y, and by $W_{XY}$ the total time spent in contact by any individual of class X with any individual of class Y. The (X,Y) entry is a bimodal distribution that assumes the value $W_{XY}/E_{XY}$ (i.e., the average observed weights between classes X and Y) with probability $E_{XY}/(N_X N_Y)$, and the value 0 with probability 1- $E_{XY}/(N_X N_Y)$. On building a network representation from the matrix, for each pair of individuals *(x,y)*, with *x* in class X and *y* in class Y, we accordingly set $w_{xy}= W_{XY}/E_{XY}$ with probability $E_{XY}/(N_X N_Y)$, and $w_{xy}= 0$ with probability 1- $E_{XY}/(N_X N_Y)$. This representation is discussed in the Additional file 1.

**Infectious disease model**

To simulate the spread of an epidemic we use an SEIR model without births, deaths or introductions of individuals. In this model, individuals can be in one of 4 distinct states: susceptible (S), exposed (E), infectious (I) or recovered (R). Susceptible individuals can become exposed with a given rate β, when they are in contact with infectious individuals. Exposed individuals do not transmit the disease through contacts with others, but become infectious with a rate σ, where 1/σ is the average duration of the latent period of the disease. Infectious individuals can transmit the disease through contacts with susceptible ones. Finally, infectious individuals become recovered with rate ν and acquire permanent immunity to the disease. The initial conditions we use correspond to a population of susceptible individuals and a single infectious individual chosen at random.

The transitions from the exposed to the infectious state and from the infectious to the recovered state do not depend on the contact behavior of each individual. On the other hand, the probability that an infectious individual transmits the disease to a susceptible one does depend on the duration of their contacts. In order to compare the simulated spreading dynamics of an epidemic for different representations of the contact patterns we need to adequately define the rate of infection β for an infectious-susceptible pair, so that in all cases we have the same average infection probability. In particular, care must be taken in comparing time-varying network representations with static network representations such as HOM or HET. The corresponding procedure is detailed in (48) and in the Supporting Text (Additional file 1).

We consider two different scenarios for the simulation of epidemic spread: in the first scenario, the latent period is set to 2 days and the average infectious period is 1 day, whereas in the second scenario we halve these intervals in order to better probe the interplay with the fast temporal scale of human contacts. The parameters we use are:
(i) scenario 1: $1/\sigma=1$ day, $1/\nu=2$ days and $\beta= 6.9 \; 10^{-4}$ $s^{-1}$ (short latent and infectious periods)
(ii) scenario 2: $1/\sigma=0.5$ day, $1/\nu=1$ day and $\beta= 2.8 \; 10^{-3}$ $s^{-1}$ (very short latent and infectious periods).

For each scenario and each data representation (DYN, HOM, HET), we performed 16,000 simulation runs of the epidemic spread. In the CMD and CMB cases we also performed 16,000 runs, obtained by generating 1,600 different contact network samples from the fitted contact matrices of distributions, and, for each of them, simulating 10 realizations of the epidemic spread.

In the following we will keep the above "microscopic" epidemiological parameters fixed, and we will simulate an SEIR model over different representations of contact patterns. An alternative approach to carry out this comparison consists in tuning the model parameters so that the spread on the different contact pattern representations leads to the same basic reproductive number $R_0$. This procedure is considered in the Additional file 1 and yields results that are similar to the ones reported below.

**Analysis**

The empirical contact data are analyzed and characterized in Ref. (33). In order for the present paper to be self-contained, we report here the contact matrix measured in the hospital ward as well as the newly introduced contact matrix of distributions (CMD).

The simulated properties of epidemic spread, based on different contact pattern representations, are quantified and compared by means of several indicators that we compute over an ensemble of stochastic realizations of the epidemic process. We focus in particular on the number of recovered individuals at the end of the spreading process: its distribution provides important information on the expected magnitude of the epidemics and can hence be used to evaluate the effectiveness of interventions. We study the impact on this distribution of the choice of the initial seed (initial infectious individual) and compute the distributions of attack rates for each role class, as such measures can be used to design and evaluate containment policies targeting specific roles or groups of individuals.

Finally, we compute the extinction probability of the epidemics, as well as the fraction of stochastic realizations that yield a global attack rate lower than 10%.

For completeness, in the Additional file 1 we also consider the peak time and the duration of the epidemic. We report as well results on the expected number of secondary infections from an initially infected individual in a completely susceptible host population (i.e., the basic reproduction number [25]).

In all cases, the results of simulations are compared with the spread simulated on the DYN representation, i.e., the representation that incorporates the most detailed information on the precise timing and duration of contacts. For our purposes, the faithfulness of a data representation is measured by how close the outcomes of the epidemics spread based on that representation are to those of the simulations based on the DYN representation.

**Results**

The dataset we use provides information on approximately 16,000 contact events, recorded during one week, among 119 individuals: 37 patients (31.1% of the total number of individuals), 31 caregivers (26.1%), 20 physicians (16.8%), 21 nurses (17.6%), and 10 ward assistants (8.4%).
Table 2 reports the contact matrix defined on role classes: the element at row X and column Y is the average over all pairs of individuals x in X and y in Y of the total time that x has spent in contact with y during the measurement week, normalized to obtain daily contact durations.
Table 3 reports the parameters of the contact matrix of distributions introduced above. As shown in the Additional file 1, we fit a negative binomial distribution to each empirical distribution of cumulated contact durations between members of classes X and Y and set the (X,Y) entry of the matrix to the values of the two fitted parameters. The negative binomial form allows the variance of the distribution to be much larger than its average, thus capturing in a simple way the breadth of the empirical distribution. We remark that no zero-boosting of the negative binomial distributions was necessary in order to reproduce the fraction of missing links from the members of class X and class Y. In other words, the fraction of pairs of individuals x in X and y in Y for which no contact was recorded (cumulated contact durations equal to 0) is naturally reproduced by the simple form of the negative binomial.

In the following we systematically discuss the results corresponding to the second set of parameter values for the SEIR model (very short latent and infectious periods) and, when appropriate, point to differences with the results from the first parameter set, which are reported in the Additional file 1. Moreover, since the FULL case corresponds to a very crude assumption of global homogeneous mixing that yields a strong overestimation of the epidemic size, we include the corresponding results in the Additional file 1 only.

Figure 2 and Table 4 summarize the distributions of the final number of cases for the various contact pattern representations.
When the epidemic reaches more than 10% of the population, the distributions of its final size, shown in Fig. 2A, can be approximately grouped in two classes. The distributions for the DYN, HET and CMD cases overlap (and they are also very similar to one another for the other parameter set, see Additional file 1). Conversely, the distributions for the HOM and CM cases display a large overestimation of the final epidemic size, with the CM case yielding a higher number of cases than the HOM one.

The probability of extinction and the fraction of stochastic runs that result in an attack rate lower than 10% (Table 4) are much smaller in the HOM and CM cases. The HET and CMD cases are comparable with one another and both are still lower than the DYN case, but closer to it (remember that DYN is the most informative representation of all). The difference between the DYN and the HET/CMD cases is smaller for the other parameter set (see Additional file 1), while even in that case HOM and CM lead to a strong underestimation of the extinction probability and of the fraction of runs with an attack rate lower than 10%. Spreading simulations based on the CMB representation (see Additional file 1) have an intermediate performance but still provide a rather bad approximation of the spreading patterns obtained from the heterogeneous contact network (HET) and from the CMD representation, which do include more information about the statistical heterogeneity of contact durations.

Table 4 also shows that the role class of the seed individual strongly influences the probability of extinction and of obtaining an attack rate lower than 10%. For larger epidemics, however, the choice of the seed does not have a strong impact on the final epidemic size, as seen in Figure 2B. For all contact representations, epidemics starting from a ward assistant are the most likely to spread. On the other hand, epidemics starting from a patient, a caregiver or a physician have a systematically larger probability of affecting only a small fraction of the population, except for the CM case in which this probability is strongly underestimated and comparable to the case in which the seed is a nurse.

Figure 3 shows the distribution of the fraction of individuals of each role class reached by the epidemic in the various scenarios. The DYN, HET and CMD representations give very similar results for both sets of parameter values: assistants and nurses are particularly affected, while doctors, patients and caregivers have smaller attack rates. On the other hand, for the HOM and CM representations the distributions are very different, with a strong global overestimation of the attack rates in every class and an incorrect ranking of the risk probabilities for the various role classes. In particular, in the CM case the attack rate for patients and caregivers is as high as that for nurses and much higher than the one for doctors, in contrast to the reference case of the DYN representation.

Finally, Figure 4 and Table 5 report the temporal evolution of the number of infectious and recovered individuals for the various contact pattern representations. The curves show the simulated values averaged over the stochastic realizations that result in an attack rate higher than 10%. The epidemic peak is higher for the HOM and CM cases and very similar for the DYN, HET and CMD representations. Consistently with the findings of Ref. (48), instead, the time of the epidemic peak only slightly depends on the specific representation (see Additional file 1).

The above comparison assumes that the epidemic parameters of the SEIR model are known and that their values are characteristic of the "microscopic" dynamics of the infectious disease, at the level of individual face-to-face contacts. As discussed in the Methods section, an alternative approach to carry out this comparison is to tune the model parameters for each case (contact pattern representation) so that all cases yield, in simulation, the same average number of secondary infections from an initially infected individual. The above comparison can then be repeated. In the Additional file 1 we show that the results reported above stay qualitatively similar.

**Discussion**

We have introduced and validated in a case study a novel way of summarizing and using in simulation high-resolution contact data that describe the interactions between individuals in a structured population. We based our work on a high-resolution time-resolved dataset on the face-to-face contacts of individuals in a hospital ward. We have defined several representations of the contact data that correspond to different levels of aggregation, from the high-resolution dynamical information on all individual contacts (DYN), passing through several intermediate representation that include individual-based static networks (HOM and HET) and contact matrices (CM, CMB and CMD), to the coarse homogeneous mixing description (FULL). In particular, we have defined a new representation in terms of a contact matrix of negative binomial distributions (CMD) that has the simplicity of a contact matrix formulation but takes into account the heterogeneity of contact durations between individuals.

To compare these representations and assess their ability to correctly inform public health policies, such as targeted vaccination strategies, we used each of the representations to simulate the spread of an infectious disease, modeled by an SEIR model, and we compared the results with the outcome of simulations based on the most detailed representation (DYN), which we regard as a gold standard.

The possibility to use a certain data representation depends on the data at hand, and each representation has advantages and limitations. In particular, both dynamical and heterogeneous network representations (DYN and HET, respectively) rely on a rather large amount of information on contact patterns. Although emerging pervasive technologies allow to gather such information in diverse contexts (4, 5, 7, 8), the data currently available remain scarce and there are no standard modeling procedures to generalize contact measurements concerning a specific population during a specific period to different periods or to larger populations. In this respect, contact matrix representations are very appealing in the case of structured populations. This is precisely the case of the data used here, in which individuals are grouped according to their known role in the hospital ward. The usual contact matrix representation gives, for each pair of individuals, an average contact rate that only depends on the respective roles of those individuals. Therefore, it can be computed for one studied population and easily generalized to larger ones (under the condition that the set of roles stays similar). However, the customary contact matrix representation suffers from an important limitation: it only retains information on the *average* contact durations of different role pairs, discarding all the heterogeneities that are known to exist in the contact properties of individuals that belong to given role classes (33).

These considerations have prompted us to move beyond the customary contact matrix representation, in particular to take into account three important features of the empirical data: 1) cumulative contact durations are known to display important temporal and role-related heterogeneities, 2) empirical contact networks are sparse, and 3) the density of links in the sub-networks restricted to given role pairs strongly depends on the specific role pair (33). We have therefore introduced a novel data representation based on a matrix of contact duration distributions. For each role pair, we have fitted the distribution of the measured cumulative contact durations using a negative binomial distribution. Interestingly, no zero-boosting was needed to reproduce the fraction of links with zero weights, i.e., the sparseness of the contact sub-networks that relate one role class to another. This distributional representation thus aggregates the individuals into role classes, but takes into account the contact heterogeneities and the sparseness of the contact network. At the same time, it retains the simplicity and versatility of the usual contact matrix representation, as it can be easily used to construct the contact patterns for a different or larger population.

We validated the usefulness of the proposed contact matrix representation using data from a real-world environment, namely a high-resolution record of the contacts within the structured population of a hospital ward. We described these high-resolution contact patterns using both the standard contact matrix representation and the representation we introduced here. We then used the two representation to simulate the epidemic spread within the hospital community. Each representation was evaluated by comparing the final size of the epidemics, the impact of the seed role class, and the final attack rates in each role class against simulations based on the much richer network-based representations (HET and DYN), which retain several heterogeneities and correlations of the empirical dataset. Our results show that the usual contact matrix representation leads to an overall strong overestimation of the outbreak probability and of the attack rate. Moreover, the relative risks of individuals belonging to different classes are not correctly estimated from the simulations performed with the usual contact matrix. On the other hand, the contact matrix of distributions (CMD) affords a very good qualitative and quantitative agreement with the spreading patterns obtained with the heterogeneous network representation (HET), even though the CMD representation contains significantly less information on the contact network structure than HET (see Table 1). The agreement is significantly better for the CMD representation than for CMB, as CMD better takes into account the heterogeneity of contact durations. The CMD representation, therefore, provides an important improvement with respect to the usual contact matrix description, as it allows to correctly rank the groups of individuals according to their risk of being infected, and therefore to correctly inform policies based on targeted vaccination of role classes.

We also note that the spreading dynamics over the (static) network representation with heterogeneous weights (HET) yields here a smaller probability of extinction than observed for the full dynamical contact data (DYN). When restricting the comparison to outbreaks with final attack rates higher than 10%, a small but systematic difference in the number of observed cases is observed, while similar timings are obtained. These results are somewhat different from the ones obtained in Ref (48) for the case of contacts at a conference. This difference might stem from various causes: first, conference attendees mix more than individuals in a hospital and the structure of the corresponding contact network is thus different from the hospital case (33). Moreover, the present hospital dataset is longer than the 2-days dataset used in Ref (48), hence dynamical effects at multiple timescales may play a stronger role. The difference between the simulations performed on the HET and DYN representations are larger for faster propagations, consistently with the results of Ref. (49), according to which a very fast (deterministic) spreading phenomenon can result in very different patterns on static and dynamic network representations.
In both cases (HET and DYN), the role class of the initial seed has a strong impact on the extinction probability and on the probability of observing a large outbreak: if the seed is a ward assistant or a nurse, the probability of a large outbreak is much larger. In addition, assistants and nurses have an overall larger risk compared to the other role classes. These results are consistent with literature that highlights the crucial importance of prioritizing nurses for local infection control interventions (35).

In our analysis we have only considered two extreme cases of temporal aggregation of the data. The DYN representation retains the full temporal information, i.e., the precise starting and ending time of each contact event. Conversely, the static network and contact matrix representations aggregate the contact behavior of individuals over a whole week. It is clearly possible to consider intermediate temporal aggregation levels that include some information on the activity schedules, for instance at the daily scale. We have thus built daily contact networks, as well as daily contact matrices and daily contact matrices of distributions: these representations take into account the fact that not all individuals are present every day, and that the contact patterns can vary from one day to the next. The results of SEIR spreading simulations, presented in the Additional file 1, are interestingly very close to the ones of representations aggregated at a weekly timescale, showing that weekly averaged data contain substantially enough information to correctly evaluate mitigation and containment measures.

We finally note that the study of time-resolved networks is still in its infancy (50). Recent studies [48,49,51,52,53] have investigated models of epidemic spread on temporal networks and studied the role of the networks' temporal properties, sometimes with contrasting results, as discussed in Ref. [54]. The comparison is however difficult, because different datasets, different epidemic models, and different parameters values have been used [54]. In particular, many studies have focused on processes that unfold rather fast with respect to the temporal scale of the network dynamics, so that the precise ordering of individual interactions is important, as well as the distribution of time intervals between contacts. In our case, conversely, the timescales of the spreading process and of the network dynamics are well separated, which explains why static representations yield results that are similar to the time-resolved representations like DYN. More systematic investigations on the interplay between the involved timescales, as well as studies using different epidemic models on the same datasets would be of strong interest for future research [54].

**Conclusions**

We have shown in a case study that the usual contact matrix representation of empirical contact patterns in a structured population can fail to correctly inform models of epidemic spreading and of the ensuing relative risk for groups. We have introduced an alternative representation of the empirical contact data based on a contact matrix of negative binomial distributions. This representation affords a tradeoff between too coarse and too detailed representations of the data. We have shown in a specific case that it allows to correctly model important features of epidemic spread, and in particular to estimate the relative risk of individuals in different role classes, while simultaneously maintaining a compact form that retains very little information from the time-varying contact network it summarizes. This is a step in the direction pointed by Ref. (44), as this representation provides a synopsis that combines simplicity, compactness of representation and modeling power, all of which are essential features for guiding decision making in public health contexts. These results provide insight about the level of detail that should be sought when performing measurement campaigns aimed at informing epidemic models and at understanding the efficacy of interventions. On the one hand, contact matrices containing only average quantities are not refined enough, and it is crucial to measure the heterogeneities of contact durations between individuals of given classes; on the other hand, the exact knowledge of who has been in contact with whom, over time, and of the precise structure of the contact network yields a description that is unnecessarily detailed and complex to manage for most practical cases.

Some limitations of the present study need to be discussed explicitly. First of all, the study was performed based on data for a relatively small population (nearly one hundred individuals), taken at one point in time, and considered diseases with short latent periods. On studying the DYN case, the long but limited temporal span of the data set required the generation of artificially extended activity timelines, obtained by repeating the empirical timeline. The static representations we considered do not take into account behavioral variations across days of the week, nor the heterogeneous amount of time spent by the various individuals in the ward. It is for instance possible that larger differences between the results of the simulations on the CMD and HET representations would be observed in larger populations because of specific network effects.

These limitations point to natural ways of extending the work reported here. In order to validate our approach in other contexts and for different categories of diseases, e.g. with longer latent periods, it is crucially important to measure high-resolution contact patterns in different structured populations, in larger populations, and over longer time scales. Moreover it would be important to check the robustness of the fitting procedure based on negative binomial distributions, and to test other possible functional forms that can fit the empirical distribution.

Other perspectives include testing data-driven containment strategies based on the various contact patterns representations and understanding how the evaluation of their performance depends on the representation, in particular for the case of the novel representation defined by the contact matrix of distributions we introduced here. A particularly interesting validation would be to extract a contact matrix of distributions from one set of data, e.g. in a hospital, to use it to design efficient containment strategies, and then to apply these strategies to a simulated epidemic on another dataset for a similar environment, for instance another hospital.

**Acknowledgments**

This work has been partially supported by the Italian Ministry of Health (Grant n. 1M22) to CR, AET and CC, by the French ANR (contract no. ANR-12-MONU-0018 HARMSFLU) to AB, and by the EU FET project Multiplex 317532 to AB and CC. The funders had no role in study design, data collection and analysis, decision to publish, or preparation of the manuscript. CC acknowledges a stimulating discussion with John Edmunds and inspiring discussions at the Data Science and Epidemiology workshop organized by Marcel Salathé.

**Author Contributions**

AB and CC designed the study. AB, CC, FG, CR, AET designed and executed the contact network measurements in the hospital. FG and AET managed the field operations, collected participant metadata and cleaned the data. AB, CC, AM analyzed the data. AM performed the simulations. AB, CC, AM wrote the manuscript. All Authors contributed to the final version of the manuscript.

**Competing Interests**

The authors declare no competing interest.

**Figure captions**

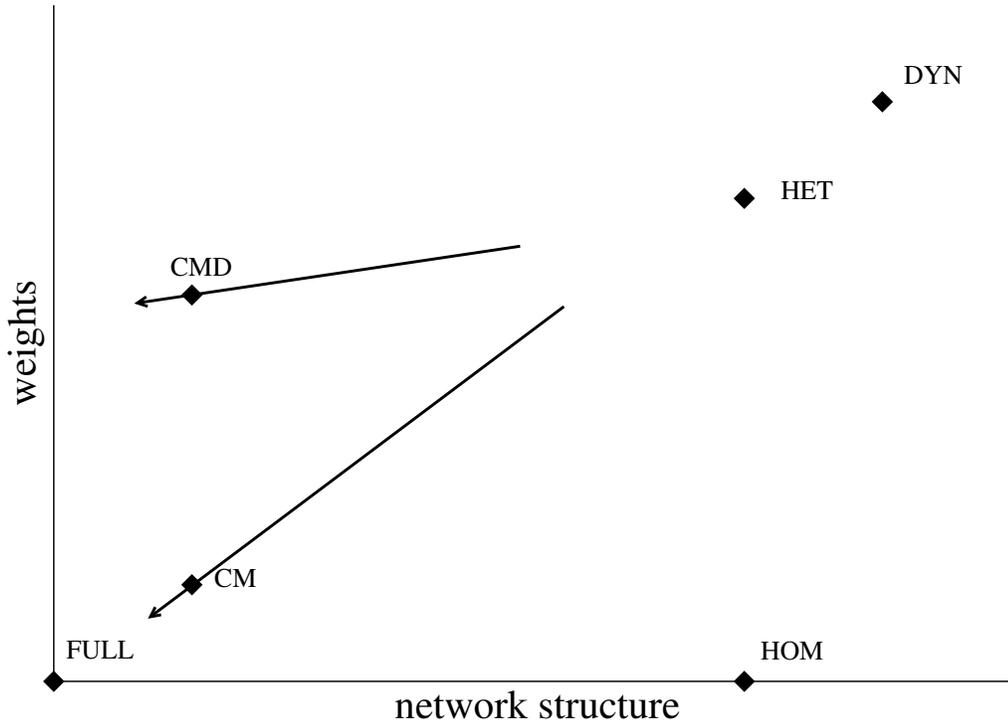

**Figure 1.** Schematic view of the considered contact patterns representations according to the two qualitative features of network detail and amount of information on contacts.

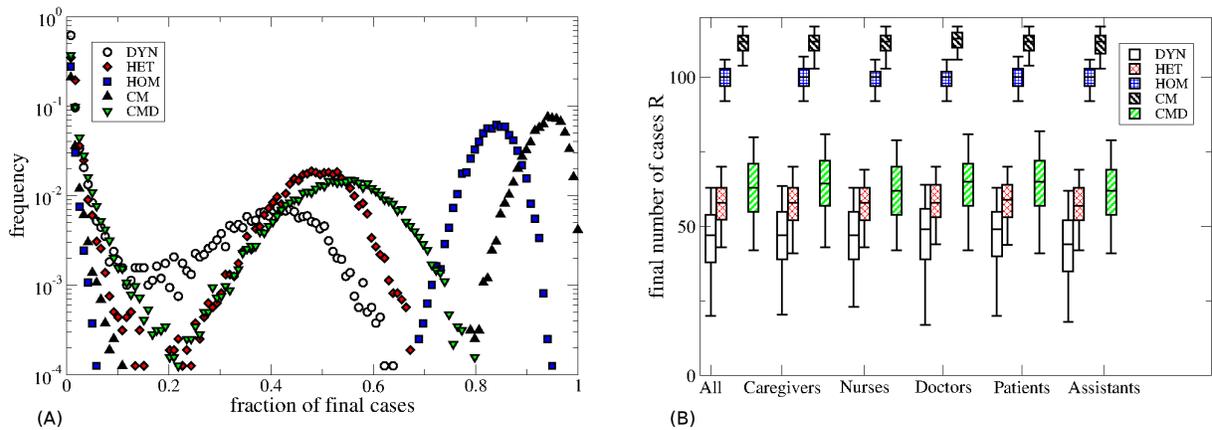

**Figure 2.** A) Distribution of the final number of cases for the various contact pattern representations, averaged over all seeds, with the parameters $1/\sigma=0.5$ days, $1/\nu=1$ day and $\beta= 2.8 \cdot 10^{-3}$ s$^{-1}$. B) Boxplots giving the distribution of the final number of cases for the various contact patterns representations, averaged over all seeds ("all") and as a function of the class of the seed, for $1/\sigma=0.5$ days, $1/\nu=1$ day and $\beta= 2.8 \cdot 10^{-3}$ s$^{-1}$. Only epidemics with a final attack rate larger than 10% are considered. The bottom and top of the rectangular boxes correspond to the 25th and 75th quantiles of the distribution, the horizontal lines to the median, and the whiskers indicate the 5th and 95th percentiles.

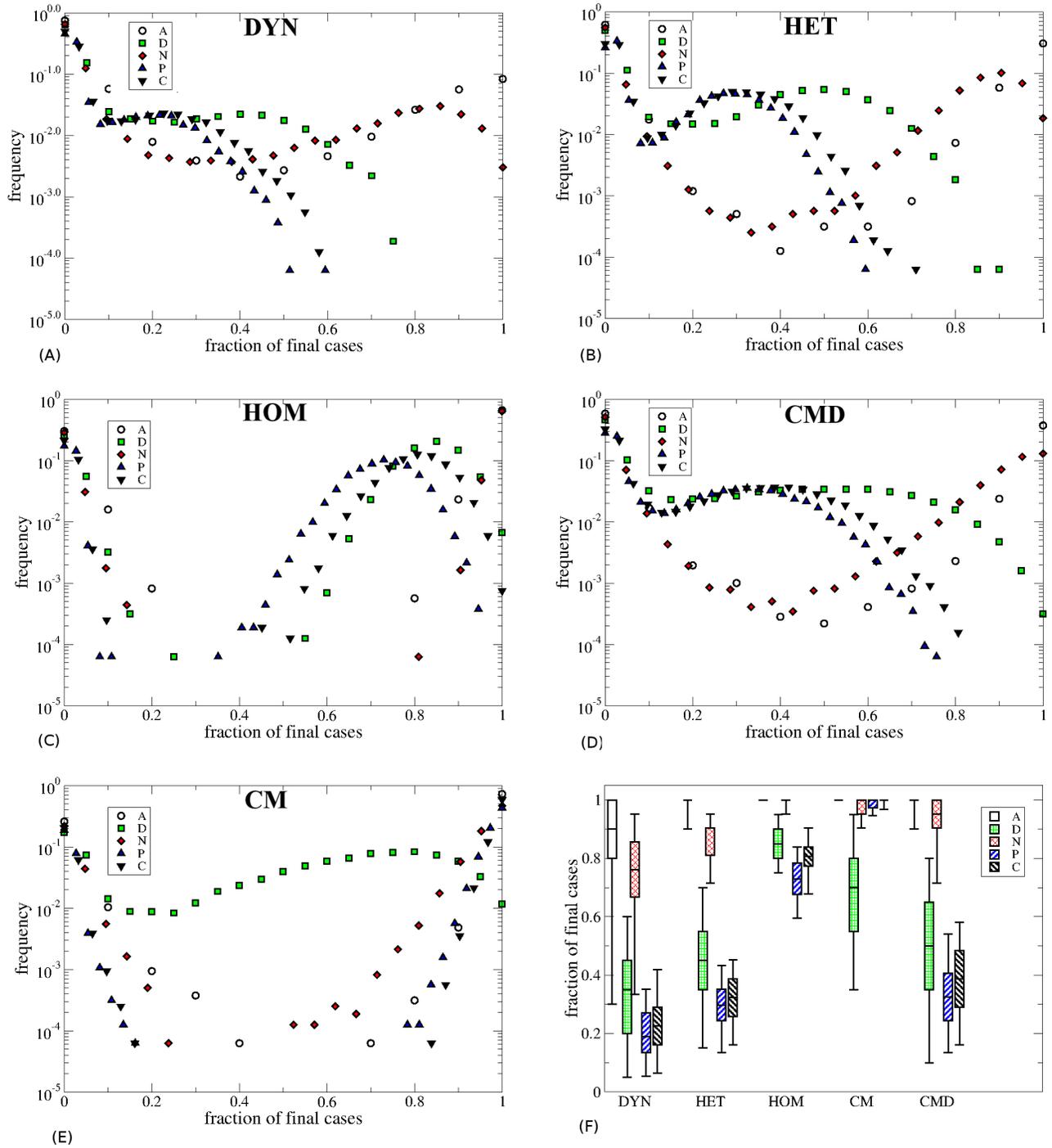

**Figure 3.** A-E) Distributions of the fraction of individuals of each class reached by the spread, for the various contact pattern representations. F) Boxplots (boxes defined as in Fig. 3) showing these distributions when the global attack rate is larger than 10%. Here $1/\sigma=0.5$ days, $1/\nu=1$ day and $\beta= 2.8 \cdot 10^{-3}$ s$^{-1}$

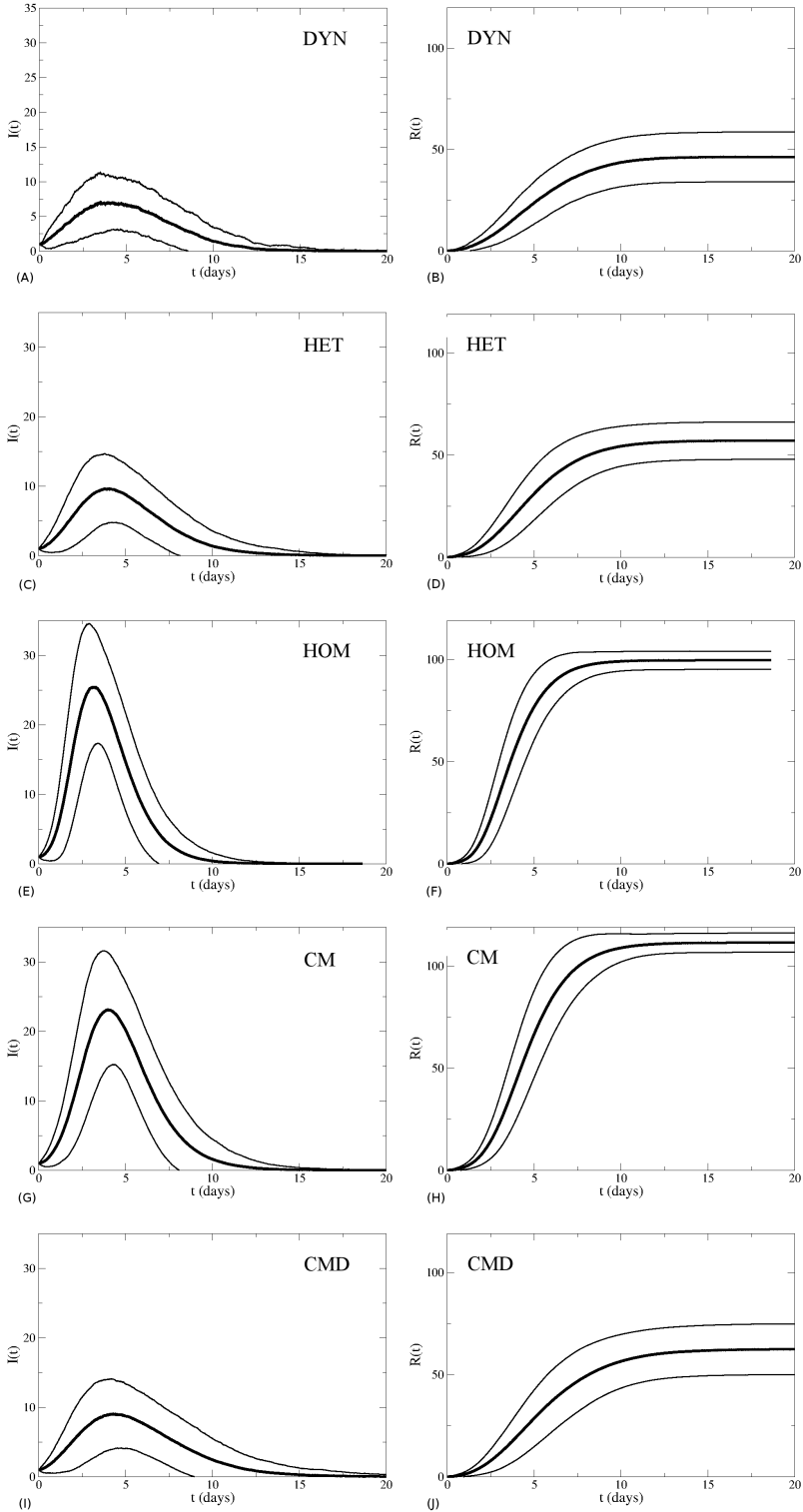

**Figure 4.** Temporal evolution of the fraction of infectious (panels A, C, E, G, I) and recovered (panels B, D, F, H, J) individuals, for the various contact pattern representations, for the second set of parameter values ($1/\sigma=0.5$ days, $1/\nu=1$ day and $\beta=2.8 \cdot 10^{-3}$ s$^{-1}$). Only runs with attack rate (AR) larger than 10% are taken into account. Bold lines represent the average values and thinner lines represent the standard deviation of the number of infectious and recovered individuals.

# Tables

**Table 1.** Summary of the properties of empirical contacts that are preserved ("X" marks) or not preserved (empty cells) by the contact patterns representations investigated here.

| DYN | HET | HOM | CM | CMD | CMB | FULL |
|---|---|---|---|---|---|---|
| Temporal ordering of contacts | | | | | | |
| Individual edge weights | X | | | | | |
| Network topology | X | X | | | | |
| Number of edges between different class pairs | X | X | | X | X | |
| Total contact time between different class pairs | X | | X | X | X | |
| Distribution of edge weights within each class pair | X | | | X | | |

**Table 2.** Contact matrix between the different roles (A: assistants; D: physicians; N: nurses; P: patients; C: caregivers). The number at row X and column Y gives the average over all pairs of individuals x in X and y in Y of the average daily time, in seconds, that x and y have spent in face-to-face proximity.

|   | A | D | N | P | C |
|---|---|---|---|---|---|
| A | 298 | 1.16 | 24.7 | 0.95 | 1.92 |
| D | 1.16 | 20.8 | 3.99 | 0.95 | 1.20 |
| N | 24.7 | 3.99 | 47.3 | 2.32 | 2.57 |
| P | 0.95 | 0.95 | 2.32 | 1.27 | 46.9 |
| C | 1.92 | 1.20 | 2.57 | 46.9 | 1.80 |

**Table 3**. Negative binomial fit parameters m and r as provided by the R's "fitdistr" function (see Additional file 1). The values at row X and column Y give the parameters obtained when fitting the empirical distribution of cumulated contact times between individuals of classes X and Y. The parameter m corresponds to the average of the fitted distribution, while its variance is given by $m+m^2/r$. The numbers in parenthesis are the standard errors as given by the "fitdistr" function.

| m parameter | A | D | N | P | C |
|---|---|---|---|---|---|
| A | 300(60) | 1.2(0.2) | 25(3) | 0.95(0.16) | 2.0(0.3) |
| D | 1.2(0.2) | 21(5) | 4.0(0.6) | 0.95(0.17) | 1.2(0.2) |
| N | 25(3) | 4.0(0.6) | 47(5) | 2.3(0.4) | 2.6(0.4) |
| P | 0.95(0.16) | 0.95(0.17) | 2.3(0.4) | 1.3(0.7) | 47(16) |
| C | 2.0(0.3) | 1.2(0.2) | 2.6(0.4) | 47(16) | 1.8(0.9) |

| r parameter | A | D | N | P | C |
|---|---|---|---|---|---|
| A | 0.615(0.014) | 0.195(0.002) | 0.404(0.0018) | 0.136(0.0007) | 0.215(0.0013) |
| D | 0.195(0.002) | 0.112($2.10^{-4}$) | 0.1278($2.10^{-4}$) | 0.0482($5.10^{-5}$) | 0.0602($8.10^{-5}$) |
| N | 0.404(0.0018) | 0.1278($2.10^{-4}$) | 0.3696(0.0013) | 0.05652($4.10^{-5}$) | 0.0845($9.10^{-5}$) |
| P | 0.136($7.10^{-4}$) | 0.0482($5.10^{-5}$) | 0.0565($4.10^{-5}$) | 0.00489($1.8.10^{-6}$) | 0.00718($9.10^{-7}$) |
| C | 0.215(0.0013) | 0.0602($8.10^{-5}$) | 0.0845($9.10^{-5}$) | 0.00718($9.10^{-7}$) | 0.009($6.10^{-6}$) |

**Table 4.** Extinction probability and fractions of runs leading to an attack rate (AR) lower than 10%, for the various contact patterns representations and as a function of the role class of the seed. Parameter values are: $1/\sigma$=0.5 days, $1/\nu$=1 day, and $\beta$=2.8 $10^{-3}$ $s^{-1}$.

| Seed class | Number of runs | DYN Ext. Prob; AR <10% | | HET Ext. Prob; AR <10% | | HOM Ext. Prob; AR <10% | | CM Ext. Prob; AR <10% | | CMD Ext. Prob; AR <10% | |
|---|---|---|---|---|---|---|---|---|---|---|---|
| All | 16000 | 0.62 | 0.81 | 0.35 | 0.63 | 0.28 | 0.32 | 0.21 | 0.27 | 0.37 | 0.59 |
| Assistants | 1344 | 0.44 | 0.51 | 0.13 | 0.16 | 0.10 | 0.12 | 0.10 | 0.12 | 0.11 | 0.14 |
| Doctors | 2690 | 0.70 | 0.84 | 0.32 | 0.64 | 0.27 | 0.30 | 0.40 | 0.55 | 0.42 | 0.66 |
| Nurses | 2823 | 0.51 | 0.64 | 0.26 | 0.33 | 0.11 | 0.13 | 0.20 | 0.26 | 0.21 | 0.30 |
| Patients | 4975 | 0.68 | 0.90 | 0.40 | 0.79 | 0.39 | 0.44 | 0.18 | 0.23 | 0.47 | 0.74 |
| Care-givers | 4168 | 0.62 | 0.89 | 0.31 | 0.77 | 0.32 | 0.37 | 0.16 | 0.20 | 0.42 | 0.72 |

**Table 5.** Summary of the average properties of the spread with AR > 10%, for $1/\sigma$=0.5 days, $1/\nu$=1 day and $\beta$= 2.8 $10^{-3}$ $s^{-1}$. The values in parenthesis provide the standard deviations.

| Contact pattern | Final size | Peak time | End time |
|---|---|---|---|
| DYN | 45(13) | 4.4(2.2) | 10.5(3.0) |
| HET | 57(9) | 4.2(1.8) | 10.6(2.6) |
| HOM | 99(5) | 3.2(0.9) | 9.1(1.7) |
| CM | 111(4) | 4.2(1.3) | 10.8(2.2) |
| CMD | 62(13) | 4.8(2.2) | 11.8(3.1) |